\documentclass[pdflatex,sn-basic]{sn-jnl}

\usepackage{graphicx}%
\usepackage{multirow}%
\usepackage{amsmath,amssymb,amsfonts}%
\usepackage{amsthm}%
\usepackage{mathrsfs}%
\usepackage[title]{appendix}%
\usepackage{xcolor}%
\usepackage{textcomp}%
\usepackage{manyfoot}%
\usepackage{booktabs}%
\usepackage{algorithm}%
\usepackage{algorithmicx}%
\usepackage{algpseudocode}%
\usepackage{listings}%

\theoremstyle{thmstyleone}%
\theoremstyle{thmstyletwo}%

\theoremstyle{thmstylethree}%

\begin{document}

\title[Article Title]{Quantum Feature Selection for Biomedical Data Analysis}


\author[1,2]{\fnm{Hongbin} \sur{Liu}}\email{liuh20@rpi.edu}

\author[3]{\fnm{Robert} \sur{Lahmann}}\email{robert.lahmann@kipu-quantum.com}

\author[4]{\fnm{Benjamin} \sur{Campbell}}\email{campbb5@rpi.edu}

\author[4,5]{\fnm{Zhemin} \sur{Zhang}}\email{zhangz29@rpi.edu}

\author[4]{\fnm{Zhiding} \sur{Liang}}\email{liangz9@rpi.edu}

\author[1]{\fnm{Siona} \sur{Bapat}}\email{bapats@rpi.edu}

\author*[1,2,6]{\fnm{Juergen} \sur{Hahn}}\email{hahnj@rpi.edu}

\affil*[1]{\orgdiv{Department of Biomedical Engineering}, \orgname{Rensselaer Polytechnic Institute}, \orgaddress{\street{110 8th Street}, \city{Troy}, \postcode{12180}, \state{NY}, \country{USA}}}

\affil[2]{\orgdiv{Center for Biotechnology \& Interdisciplinary Studies}, \orgname{Rensselaer Polytechnic Institute}, \orgaddress{\street{110 8th Street}, \city{Troy}, \postcode{12180}, \state{NY}, \country{USA}}}

\affil[3]{\orgname{Kipu Quantum GmbH}, \orgaddress{\street{Greifswalderstrasse 212}, \city{Berlin}, \postcode{10405}, \country{Germany}}}

\affil[4]{\orgdiv{Department of Computer Science}, \orgname{Rensselaer Polytechnic Institute}, \orgaddress{\street{110 8th Street}, \city{Troy}, \postcode{12180}, \state{NY}, \country{USA}}}

\affil[5]{\orgdiv{Department of Electrical, Computer, and Systems Engineering}, \orgname{Rensselaer Polytechnic Institute}, \orgaddress{\street{110 8th Street}, \city{Troy}, \postcode{12180}, \state{NY}, \country{USA}}}

\affil[6]{\orgdiv{Department of Chemical \& Biological Engineering}, \orgname{Rensselaer Polytechnic Institute}, \orgaddress{\street{110 8th Street}, \city{Troy}, \postcode{12180}, \state{NY}, \country{USA}}}



\abstract{Feature selection is an essential step for reducing complexity of high dimensional data, usually in preparation for developing machine learning models such as computational biomarkers. However, there are limitations associated with feature selection such as for metabolomic data where there are hundreds or even thousands of features per study participant, while the available number of participants in a clinical trials is limited. Classical methods such as exhaustive search require evaluation of all possible feature combinations, making them costly in terms of computation and runtime, or even infeasible, as feature dimensionality increases. In this study, we propose a novel Quadratic Unconstrained Binary Optimization (QUBO) coefficient formulation and pose metabolomic feature selection as a QUBO problem that selects a specified number of features by balancing their relevance against redundancy among the selected variables. To evaluate our proposed QUBO objective function, we conducted a series of experiments using Bias-Field Digitized Counterdiabatic Quantum Optimization (BF-DCQO) and Quantum Approximate Optimization Algorithm (QAOA) on a quantum gate based computer. We also compared the method to several classical methods on three metabolomic datasets associated with Autism Spectrum Disorder (ASD) on a classical computer. Our method reduces runtime compared with exhaustive search and Iterative Tabu Search (ITS) and achieves competitive performance across classifiers compared to classical filter, wrapper, and embedded methods. These results do not claim quantum advantage; rather, they establish hardware feasibility and demonstrate the current capabilities of Noisy Intermediate-Scale Quantum (NISQ) devices.}


\maketitle

\section{Introduction}\label{sec1}

Autism Spectrum Disorder (ASD) is a neuro-developmental condition characterized by differences in social communication and interaction, together with restricted or repetitive patterns of behavior and interests \citep{Hodges2020}. In 2022, the estimated prevalence of ASD was 32.2 per 1,000, or approximately one in 31, among 8-year-old children in the United States \citep{Shaw2025}. ASD is thought to arise from interactions of genetic and environmental factors, with multiple etiological pathways rather than a single unifying cause \citep{Szatmari2003}. Furthermore, while it is considered ideal to diagnose autism at 18-24 months, the average age for an ASD diagnosis is approximately 4 years, leaving a 2-2.5 year gap where no intervention was performed due to the lack of a diagnosis \citep{ADDM2014, McCarty2020}. Challenges related to the diversity of presentation of autism as well as diagnosing young children has motivated biomarker-based approaches to emerge as a promising means of earlier and potentially more precise ASD screening and diagnostic evaluation \citep{Flynn2026, AlSaei2024, Vargason2020}. However, biomarker-based approaches such as metabolomic measurements can generate hundreds or even thousands of features per study participant, while the number of participants in a clinical trial is limited. This imbalance between feature dimensionality and sample size creates a high-dimensional classification problem in which models can easily overfit the data. As a result, identifying small and stable subsets of features is essential for building reliable classification models.

Feature selection is a strategy for addressing high dimensional problems by selecting a subset of the features without transforming them into new representations. Removing irrelevant and redundant features can preserve discriminative information while reducing model complexity and improve generalization \citep{Chandrashekar2014}. These benefits are important in metabolomic studies, where many correlated pathway measurements and limited sample sizes can increase the risk of overfitting. Feature selection can therefore reduce the number of variables to produce more stable and predictive models while improving interpretability by identifying the metabolites that contribute most strongly to classification. The objective of this study is to address the feature selection problem on a metabolomic data set for classification. Classical feature selection methods are generally categorized as filter, wrapper, and embedded approaches \citep{Chandrashekar2014}. Among these approaches, exhaustive search \citep{Trakhtenbrot1984} and greedy algorithm \citep{Edmonds1971} are commonly used wrapper methods in metabolomic studies. Exhaustive search evaluates every possible feature subset and guarantees the optimal solution, whereas greedy algorithm iteratively add features; both use a classifier as a black box and optimize its predictive performance. However, the number of possible combinations increases rapidly for feature selection problems as the total number of candidate features or the desired subset size grows, making exhaustive search computationally infeasible for high-dimensional datasets while greedy algorithms are sub-optimal. These limitations motivate global optimization methods capable of exploring large feature spaces more efficiently.

Quantum computing has emerged as a potential method for addressing complex problems in biomedical studies \citep{Repetto2025, Zheng2026, Liu2026}. One promising application is quantum feature selection formulated as a Quadratic Unconstrained Binary Optimization (QUBO) problem \citep{Muecke2023, Kochenberger2014}. In this formulation, the linear and quadratic terms encode feature relevance and redundancy, and produce a binary feature mask that represents whether individual features are selected. The resulting objective can be mapped to an Ising Hamiltonian and evaluated using quantum optimization algorithms on gate-based quantum computers. Gate-based approaches include the Quantum Approximate Optimization Algorithm (QAOA) \citep{Farhi2014}, which uses a hybrid variational procedure to search for low-energy solutions, and Bias-Field Digitized Counterdiabatic Quantum Optimization (BF-DCQO) \citep{GomezCadavid2025}, which combines digitized counterdiabatic evolution with iteratively updated bias fields to guide the system toward favorable solutions. Previous studies have demonstrated the feasibility of binary optimization based feature selection on quantum computers, including both QUBO and higher order formulations, supporting its potential as a global optimization method for identifying informative feature subsets \citep{Muecke2023, Romero2025, Pranjic2026, FloresGarrigos2026}.

The contributions of this study are twofold. First, we propose a QUBO objective function that uses Fisher score \citep{Fisher1936} to calculate feature importance and covariance to penalize pairwise feature redundancy. Second, we benchmark this formulation on three metabolomic datasets related to ASD against Mutual Information Quadratic Unconstrained Binary Optimization (MIQUBO) \citep{Muecke2023} using BF-DCQO and QAOA on gate based quantum computers, as well as against other classical feature selection methods on a classical computer, to demonstrate its feasibility.


\section{Methods and Materials}\label{sec2}

\subsection{QUBO Formulation}

QUBO feature selection represents the selection of a feature subset as a binary optimization problem in which linear terms quantify feature importance and quadratic terms penalize redundancy between selected features. Earlier work \citep{Muecke2023} proposed a QUBO framework using Mutual Information (MI) to quantify both feature relevance and pairwise redundancy. Related works also incorporated Conditional Mutual Information (CMI) to account for dependencies among features \citep{FerrariDacrema2022}. Building on this framework, we introduce a different formulation of the QUBO coefficients, using the Fisher score to measure class discriminative relevance and covariance to measure pairwise feature redundancy.

Consider a binary classification dataset $\mathcal{D}=\{(\mathbf{x}^{(k)},y^{(k)})\}_{k=1}^{N}$, where $\mathbf{x}^{(k)}\in\mathbb{R}^{n}$ contains $n$ feature measurements for subject $k$, and $y^{(k)}\in\{0,1\}$ denotes the corresponding class label. A candidate feature subset is represented by the binary indicator vector $\mathbf{x}^{*}=(x_1,\ldots,x_n)\in\{0,1\}^{n}$, where $x_i=1$ indicates that feature $i$ is selected and $x_i=0$ otherwise. The optimal feature subset is obtained by solving

\begin{equation}
\mathbf{x}^{*}
=
\underset{\mathbf{x}\in\{0,1\}^{n}}{\arg\min}
Q(\mathbf{x},\alpha),
\label{eq:qubo_opt}
\end{equation}

where the parameter $\alpha\in[0,1]$ is used both to balance the relevance and redundancy terms and to control the target number of selected features. Increasing $\alpha$ places greater weights on feature relevance and produces larger subsets, whereas decreasing $\alpha$ increases the penalty on feature redundancy. For each target feature set size, $\alpha$ was adjusted until the optimal solution contained the desired number of selected features, avoiding the need for an additional penalty term \citep{Muecke2023}.

To determine the binary feature mask $\mathbf{x}^{*}$, we formulate the objective as a balance between feature importance and pairwise redundancy. The QUBO objective function is defined as

\begin{equation}
Q(\mathbf{x},\alpha)
:=
-\alpha
\sum_{i=1}^{n}
I_i x_i
+
(1-\alpha)
\sum_{i,j=1}^{n}
R_{ij}x_i x_j,
\label{eq:qubo_objective}
\end{equation}

where $I_i$ denotes the importance of feature $i$ and $R_{ij}$ denotes the pairwise redundancy between features $i$ and $j$. The negative feature importance term favors the selection of informative features, and the positive feature redundancy term penalizes the selection of redundant features.

In our proposed formulation, feature importance term $I_i$ is calculated using the Fisher score \citep{Fisher1936} instead of the commonly used MI. For feature $i$, the Fisher score is defined as,

\begin{equation}
F_i
:=
\frac{
\left(\mu_{i,1}-\mu_{i,0}\right)^2
}{
\sigma_{i,1}^{2}+\sigma_{i,0}^{2}
},
\label{eq:fisher_score}
\end{equation}

where $\mu_{i,c}$ and $\sigma_{i,c}^{2}$ denote the mean and variance of feature $i$ for class $c\in\{0,1\}$, respectively. A larger Fisher score indicates greater separation between the two classes and therefore greater feature importance. The importance coefficient is normalized as $I_i := F_i / \max_{k} F_k$, such that $I_i \in [0,1]$.

We calculate the pairwise feature redundancy term using the absolute covariance instead of MI,

\begin{equation}
C_{ij}
:=
\left|
\operatorname{Cov}(X_i,X_j)
\right|,
\qquad i\neq j,
\label{eq:covariance}
\end{equation}

where larger values indicate greater dependence between features $i$ and $j$. The redundancy coefficient is normalized as $R_{ij} := C_{ij} / \max_{k<l} C_{kl}$ for $i \neq j$, with $R_{ii}=0$. 

Normalizing the Fisher score and covariance terms places their weights on comparable scales. Minimizing Eq.~\ref{eq:qubo_objective} produces feature subsets that provide strong class separation while limiting the selection of redundant features.

\begin{figure}[htbp]
\centering
\includegraphics[width=\textwidth]{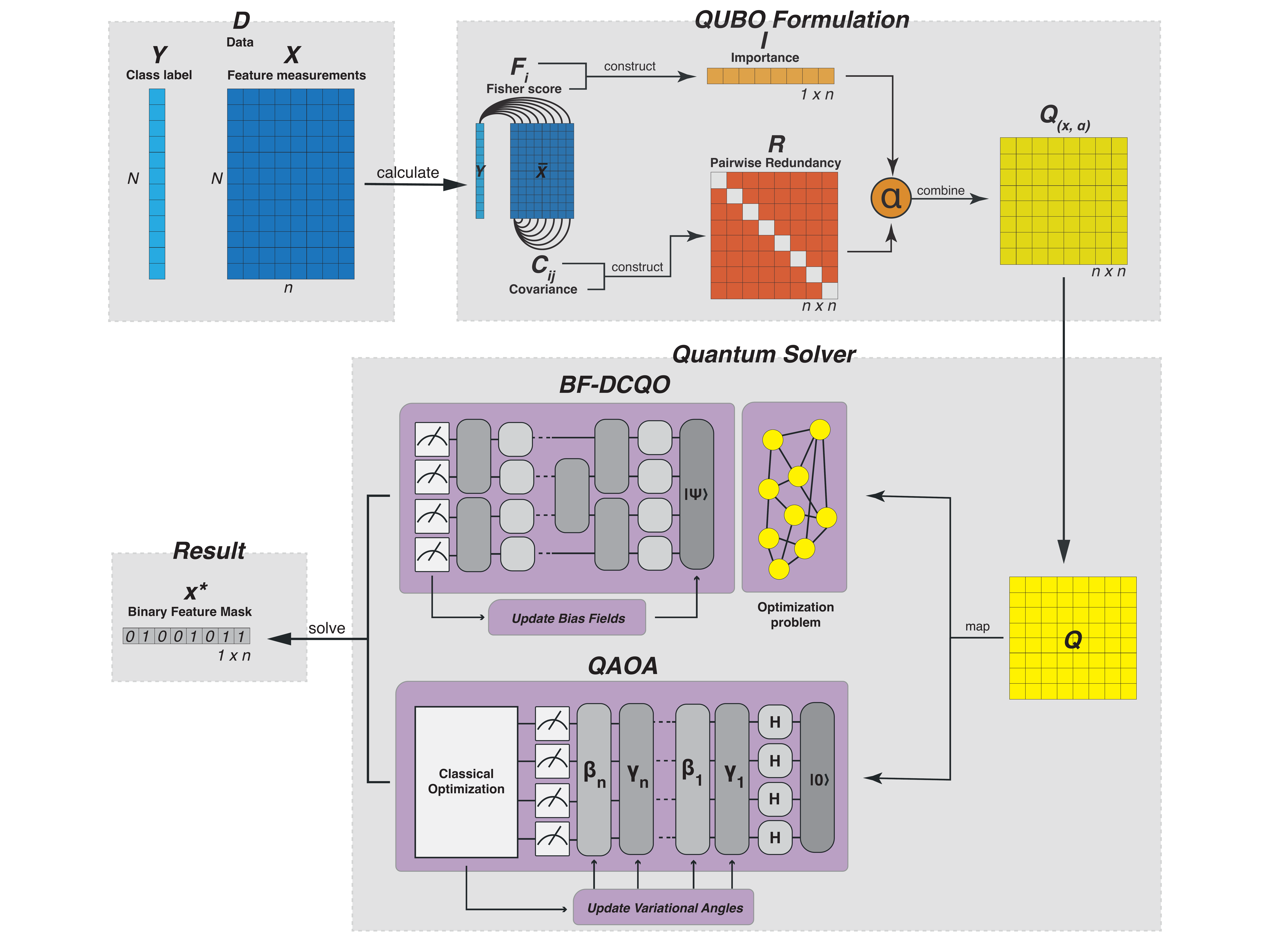}
\caption{ The quantum feature selection pipeline. From a given dataset, the Fisher scores $F_i$ and pairwise covariances $C_{ij}$ are calculated to obtain the importance vector $\mathbf{I}$ and redundancy matrix $\mathbf{R}$. These terms are combined with a parameter $\alpha$ to construct the QUBO objective $Q(\mathbf{x},\alpha)$ in Eq.~\ref{eq:qubo_objective}. The QUBO is mapped to the corresponding Ising optimization problem in Eq.~\ref{eq:ising_hamiltonian} and solved using either BF-DCQO or QAOA. The resulting binary vector $\mathbf{x}^{*}$ indicates the selected features.}
\label{fig:qfs_pipeline}
\end{figure}

For comparison, we also implemented the MIQUBO \citep{Muecke2023}. Each continuous feature was first discretized into $B=20$ quantile bins to estimate the discrete probability distributions. Then feature importance was defined as $I_i^{\mathrm{MI}}=I(X_i;Y)$, which measures the statistical dependence between feature $X_i$ and the class label $Y$, where larger values indicating greater class relevant information. Pairwise redundancy was defined as $R_{ij}^{\mathrm{MI}}=I(X_i;X_j)$, which measures the amount of information shared between features $X_i$ and $X_j$. Larger redundancy values indicate two features contain more overlapping information. A complete pipeline of the quantum feature selection workflow is shown in Fig.~\ref{fig:qfs_pipeline}.

\subsection{Quantum Optimization}

To solve the QUBO problem on a gate-based quantum computer, the binary variables are mapped to quantum operators according to
$x_i \rightarrow (\mathbb{I}-\sigma_i^{z})/2$.
The QUBO objective in Eq.~\ref{eq:qubo_objective} can then be expressed, up to an additive constant, as an Ising cost Hamiltonian

\begin{equation}
H_{\mathrm{C}}
=
\sum_{i=1}^{n} h_i \sigma_i^{z}
+
\sum_{\substack{i,j=1 \\ i\neq j}}^{n} J_{ij}\sigma_i^{z}\sigma_j^{z}
+ c,
\label{eq:ising_hamiltonian}
\end{equation}

where the coefficients $h_i$ and $J_{ij}$ come from transforming the linear feature importance and quadratic redundancy terms of the QUBO objective under the mapping $x_i \rightarrow (\mathbb{I}-\sigma_i^{z})/2$. The term $c$ is the constant energy offset introduced by the transformation and does not affect the minimizing solution. The binary variables are mapped to spin variables through $x_i=(1-\sigma_i^{z})/2$, where $\sigma_i^{z}$ is the Pauli-$Z$ operator acting on qubit $i$. The ground state of $H_{\mathrm{C}}$ therefore corresponds to the feature mask that minimizes the QUBO objective. In this study, the Ising optimization problem Hamiltonian was solved using two gate-based quantum optimization approaches: BF-DCQO and QAOA. 

BF-DCQO is a non-variational method for solving Ising optimization problems on gate  based quantum computers \citep{GomezCadavid2025}. It extends digitized counterdiabatic quantum optimization \citep{Hegade2022} by incorporating iteratively updated bias fields into the initial Hamiltonian. The evolution is described by

\begin{equation}
H(\lambda)
=
[1-\lambda(t)]\widetilde{H}_{\mathrm{i}}
+
\lambda(t)H_{\mathrm{C}}
+
\dot{\lambda}(t)A_{\lambda}^{(\ell)},
\label{eq:bfdcqo_hamiltonian}
\end{equation}

where $\lambda(t)$ controls the interpolation, $A_{\lambda}^{(\ell)}$ is an approximate adiabatic gauge potential that provides the counterdiabatic correction, and $\widetilde{H}_{\mathrm{i}}$ is the bias-modified initial Hamiltonian. In BF-DCQO, this Hamiltonian is

\begin{equation}
\widetilde{H}_{\mathrm{i}}
=
\sum_{i=1}^{n}
\left(
h_i^{x}\sigma_i^{x}
-
h_i^{b}\sigma_i^{z}
\right),
\qquad
h_i^{b}
=
\langle \sigma_i^{z} \rangle,
\label{eq:bias_hamiltonian}
\end{equation}

where $h_i^{x}$ is the transverse field strength and $h_i^{b}$ is the longitudinal bias for qubit $i$. The bias is estimated from the measured spin expectation value of the preceding iteration. The first iteration uses $h_i^{b}=0$ and subsequent iterations update the bias fields using measurements from proceeding iteration.

QAOA is a hybrid variational quantum algorithm for combinatorial optimization \citep{Farhi2014}. Starting from the uniform superposition state $|+\rangle^{\otimes n}$, QAOA alternates between evolution under the cost Hamiltonian $H_{\mathrm{C}}$ and the mixer Hamiltonian $H_{\mathrm{M}}=\sum_{i=1}^{n}\sigma_i^{x}$. For $p$ QAOA layers, the variational state is

\begin{equation}
|\psi_p(\boldsymbol{\gamma},\boldsymbol{\beta})\rangle
=
e^{-i\beta_p H_{\mathrm{M}}}
e^{-i\gamma_p H_{\mathrm{C}}}
\cdots
e^{-i\beta_1 H_{\mathrm{M}}}
e^{-i\gamma_1 H_{\mathrm{C}}}
|+\rangle^{\otimes n},
\label{eq:qaoa_state}
\end{equation}

where $\boldsymbol{\gamma}=(\gamma_1,\ldots,\gamma_p)$ and $\boldsymbol{\beta}=(\beta_1,\ldots,\beta_p)$ are variational parameters. In this study, QAOA was implemented with $p=2$ layers, giving two cost angles $(\gamma_1,\gamma_2)$ and two mixer angles $(\beta_1,\beta_2)$, for a total of four variational parameters. The variational parameters were optimized using the derivative-free Constrained Optimization By Linear Approximation (COBYLA) algorithm \citep{Powell1994}. At each objective evaluation, the QAOA circuit was sampled to estimate the expectation value $\langle H_{\mathrm{C}}\rangle$, and COBYLA used these estimates to update the variational parameters. After the classical optimization, the circuit was sampled using the optimized parameters to obtain candidate feature subsets represented by the measured bitstrings.

\begin{figure*}[htbp]
\centering
\includegraphics[width=\textwidth]{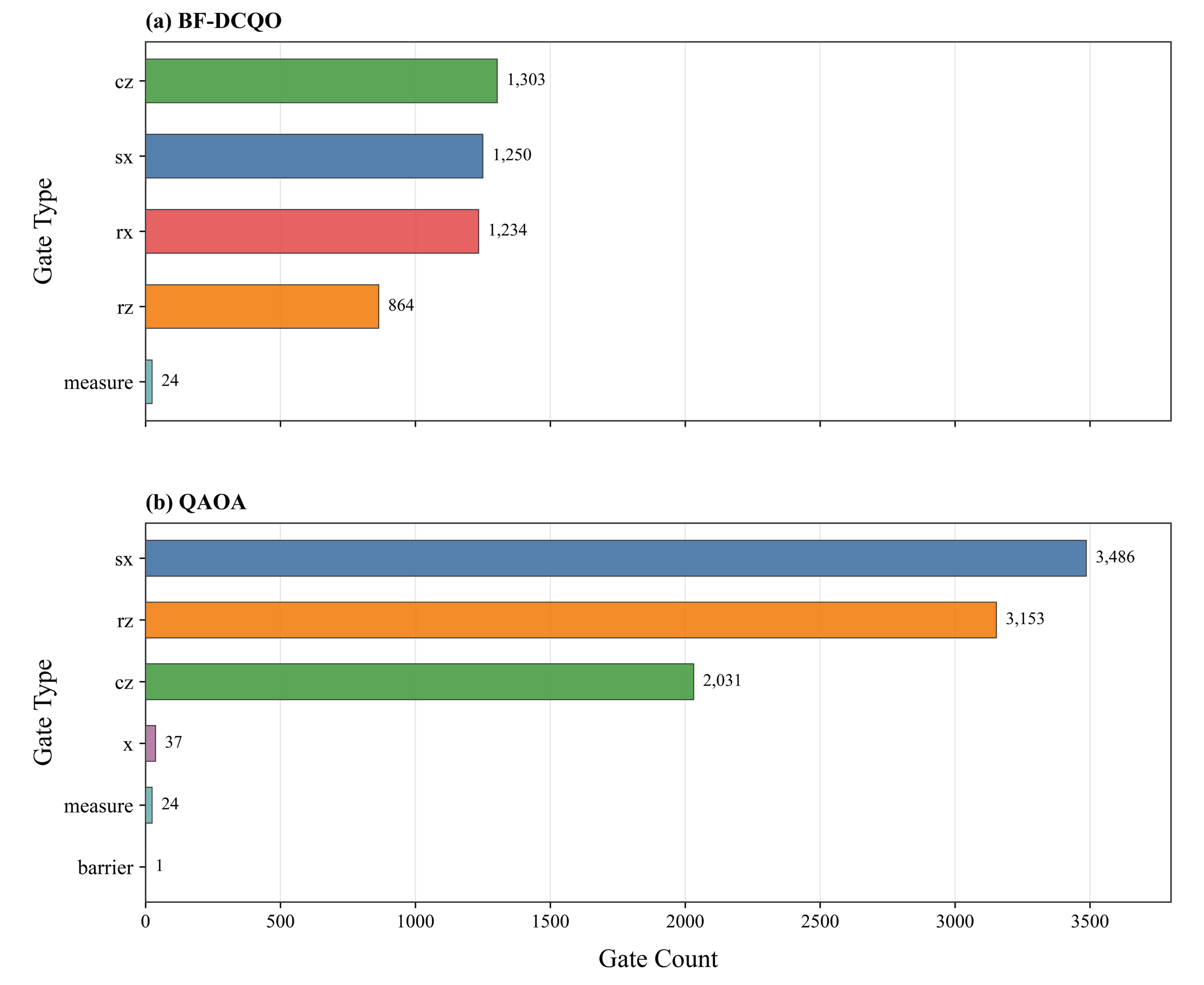}
\caption{Transpiled circuit gate counts for (a) BF-DCQO and (b) QAOA. 
For a 24 input features, BF-DCQO required 4,675 gates with a circuit depth of 777, and QAOA required 8,731 gates with a circuit depth of 2,011. 
Both circuits were mapped to 120 physical qubits with a circuit width of 144.}
\label{fig:circuit_resources}
\end{figure*}

For both BF-DCQO and QAOA, the measured bitstrings were evaluated using the QUBO objective, and the lowest QUBO energy sampled bitstring solution was taken as the selected feature subset. The transpiled circuit resources of BF-DCQO and QAOA for the 24feature QUBO instance is compared in Fig.~\ref{fig:circuit_resources}.

\subsection{Classical Baselines}

To benchmark the proposed QUBO feature selection method, we considered representative classical approaches from the three major feature selection strategies: filter, wrapper, and embedded methods \citep{Chandrashekar2014}. Additionally, we evaluated Iterated Tabu Search (ITS) as a classical optimization method for solving the QUBO objective.

Filter methods rank features using statistical measures. We evaluated Pearson correlation ranking. For feature $X_i$ and class label $Y$, the Pearson correlation score is defined as

\begin{equation}
\rho_i
=
\frac{\operatorname{Cov}(X_i,Y)}
{\sigma_{X_i}\sigma_Y},
\label{eq:pearson}
\end{equation}

where $\operatorname{Cov}(X_i,Y)$ denotes the covariance between feature $X_i$ and the class label, and $\sigma_{X_i}$ and $\sigma_Y$ denote their standard deviations. Features are ranked according to $|\rho_i|$, with larger values indicating stronger linear association with the class label \citep{Guyon2003}.

Wrapper methods evaluate candidate feature subsets directly using the predictive performance of a classifier. For a target subset size $m$, the wrapper objective can be expressed as

\begin{equation}
S^{*}
=
\underset{S\subseteq\{1,\ldots,n\},\,|S|=m}
{\arg\max}
\operatorname{Acc}_{\mathrm{SVM}}(S),
\label{eq:wrapper_objective}
\end{equation}

where $S$ denotes a candidate subset, $|S|=m$ constrains the subset to the target number of features, and $\operatorname{Acc}_{\mathrm{SVM}}(S)$ denotes the classification accuracy of a linear Support Vector Machine (SVM) trained using subset $S$. In this study, exhaustive search was used as the representative wrapper method. It evaluates all $\binom{n}{m}$ possible feature subsets and therefore identifies the subset with the highest classification accuracy under the specified SVM evaluation criterion. Its computational cost grows combinatorially with the number of candidate features and the target subset size \citep{Chandrashekar2014}.

For embedded method baseline, we evaluated Support Vector Machine Recursive Feature Elimination (SVM-RFE) \citep{Chapelle2008}. For a linear Support Vector Machine (SVM), the decision function is

\begin{equation}
f(\mathbf{x})
=
\mathbf{w}^{T}\mathbf{x}+b,
\label{eq:svm}
\end{equation}

where $\mathbf{w}$ is the learned feature weight vector and $b$ is the bias term. The magnitude of each component $w_i$ reflects the contribution of feature $i$ to the separating hyperplane \citep{Cortes1995}. SVM-RFE therefore ranks each feature according to $r_i=w_i^{2}$, where a smaller $r_i$ indicates a smaller contribution to the SVM decision function. At each iteration, the feature with the smallest ranking score is removed and the SVM is retrained until it reaches the target number of features.

In addition to the feature selection baselines, ITS was implemented as a classical solver for the proposed QUBO formulation \citep{Palubeckis2006} using the MATLAB Support Package for Quantum Computing add-on. ITS performs local search over binary solutions while maintaining a Tabu memory to restrict recently visited moves and reduce cycling. The search repeats from promising solutions to escape local optima and explore additional regions of the solution space.

\subsection{Performance Evaluation}

The feature subsets selected were evaluated using five classification models: Linear Discriminant Analysis (LDA), linear Support Vector Machines (SVM), Logistic Regression (LR), Random Forest (RF), and Partial Least Squares Discriminant Analysis (PLS-DA).

LDA classifies a sample by comparing class linear discriminant functions \citep{Fisher1936}. For class $c$, the discriminant function is

\begin{equation}
\delta_c(\mathbf{z})
=
\mathbf{z}^{T}\mathbf{\Sigma}^{-1}\boldsymbol{\mu}_c
-
\frac{1}{2}
\boldsymbol{\mu}_c^{T}\mathbf{\Sigma}^{-1}\boldsymbol{\mu}_c
+
\log \pi_c,
\label{eq:lda}
\end{equation}

where $\mathbf{z}$ is the input feature vector, $\boldsymbol{\mu}_c$ is the mean vector of class $c$, $\mathbf{\Sigma}$ is the shared within-class covariance matrix, and $\pi_c$ is the prior probability of class $c$. The predicted class is $\hat{y}=\arg\max_c \delta_c(\mathbf{z})$.

Linear SVM determines a separating hyperplane by maximizing the margin between the two classes \citep{Cortes1995}. The optimization problem is

\begin{equation}
\underset{\mathbf{w},b,\boldsymbol{\xi}}{\min}
\;
\frac{1}{2}\|\mathbf{w}\|^{2}
+
C\sum_{k=1}^{N}\xi_k,
\label{eq:svm_classifier}
\end{equation}

subject to $y_k(\mathbf{w}^{T}\mathbf{z}^{(k)}+b)\geq1-\xi_k$ and $\xi_k\geq0$, where $\mathbf{w}$ and $b$ define the separating hyperplane, $\xi_k$ are slack variables, and $C$ controls the trade-off between margin maximization and classification error. Prediction is determined by the sign of $\mathbf{w}^{T}\mathbf{z}+b$.

Logistic Regression \citep{Cox1958} models the probability that a sample belongs to the positive class as

\begin{equation}
P(Y=1\mid\mathbf{z})
=
\frac{1}
{1+\exp[-(\mathbf{w}^{T}\mathbf{z}+b)]},
\label{eq:logistic}
\end{equation}

where $\mathbf{w}$ is the learned coefficient vector and $b$ is the intercept. A sample is assigned to the positive class when the estimated probability exceeds the classification threshold.

Random Forest combines an ensemble of $B$ decision trees trained using bootstrap samples and randomly selected feature subsets \citep{Breiman2001}. The final prediction is obtained through majority voting,

\begin{equation}
\hat{y}
=
\operatorname{mode}
\left\{
T_1(\mathbf{z}),
T_2(\mathbf{z}),
\ldots,
T_B(\mathbf{z})
\right\},
\label{eq:random_forest}
\end{equation}

where $T_b(\mathbf{z})$ denotes the class prediction produced by the $b$th decision tree.

PLS-DA constructs latent components that maximize the covariance between the predictor matrix $\mathbf{Z}$ and the class response $\mathbf{y}$ \citep{Barker2003}. For latent component $h$, the score vector is

\begin{equation}
\mathbf{t}_h
=
\mathbf{Z}\mathbf{w}_h,
\qquad
\mathbf{w}_h
=
\underset{\|\mathbf{w}\|=1}{\arg\max}
\operatorname{Cov}^{2}
\left(
\mathbf{Z}\mathbf{w},\mathbf{y}
\right),
\label{eq:pls-da}
\end{equation}

where $\mathbf{w}_h$ is the weight vector defining the $h$-th latent component. The class response is then estimated from the resulting latent representation, with the predicted continuous response converted to the corresponding binary class.

Classification performance was evaluated using Leave-One-Out Cross-Validation (LOOCV) \citep{Stone1974}. For a dataset containing $N$ subjects, LOOCV consists of $N$ folds, where subject $k$ is held out for testing while the remaining $N-1$ subjects are used for model training. The process is repeated until every subject has served once as the test sample. Overall classification accuracy is calculated as

\begin{equation}
\operatorname{Acc}
=
\frac{1}{N}
\sum_{k=1}^{N}
\mathbb{I}
\left(
\hat{y}^{(k)}=y^{(k)}
\right),
\label{eq:loocv_accuracy}
\end{equation}

where $y^{(k)}$ and $\hat{y}^{(k)}$ denote the true and predicted class labels for subject $k$, respectively, and $\mathbb{I}(\cdot)$ is the indicator function. Feature standardization was performed independently within each training fold and then applied to the corresponding held-out subject to prevent information leakage. The same LOOCV procedure was applied to all feature selection methods.

\subsection{Datasets and Hardware}

Three metabolomic datasets collected for classification of autism spectrum disorder were used in this study. Dataset 1 contains plasma measurements of the folate-dependent one-carbon metabolism and the transsulfuration (FOCM/TS) pathways \citep{Melnyk2012}, Dataset 2 contains measurements resulting from untargeted plasma metabolomics \citep{Kang2020}, and Dataset 3 contains measurements of shotgun metagenomic profiles of the gut microbiome \citep{Nirmalkar2022}. Since the de-identified datasets have already been published in earlier works and this study just makes use of these de-identified datasets, no IRB review was required for this study.

Due to the limited qubit capacity and executable circuit depth of NISQ hardware, the dimensionality of the larger datasets, i.e., dataset 2 and 3, was reduced. Following the multiple hypothesis testing prescreening approach proposed by an earlier study \citep{Arici2025}, Dataset 2 and 3 were reduced to 56 and 54 features, respectively. This prescreening step reduces the quantum search space while retaining features that exhibit differences between the ASD and Typically Developing (TD) groups. A summary of the feature dimensionality and class distribution for all three datasets is provided in Table~\ref{tab:datasets}.

\begin{table}[h]
\caption{Summary of the datasets used in this study.}\label{tab:datasets}%
\begin{tabular}{@{}lccc@{}}
\toprule
Dataset & Features & ASD subjects & TD subjects \\
\midrule
Dataset 1 & 24 & 83 & 76 \\
Dataset 2 & 56 & 18 & 20 \\
Dataset 3 & 54 & 18 & 20 \\
\botrule
\end{tabular}
\end{table}

\begin{figure}[htbp]
\centering
\includegraphics[width=\textwidth]{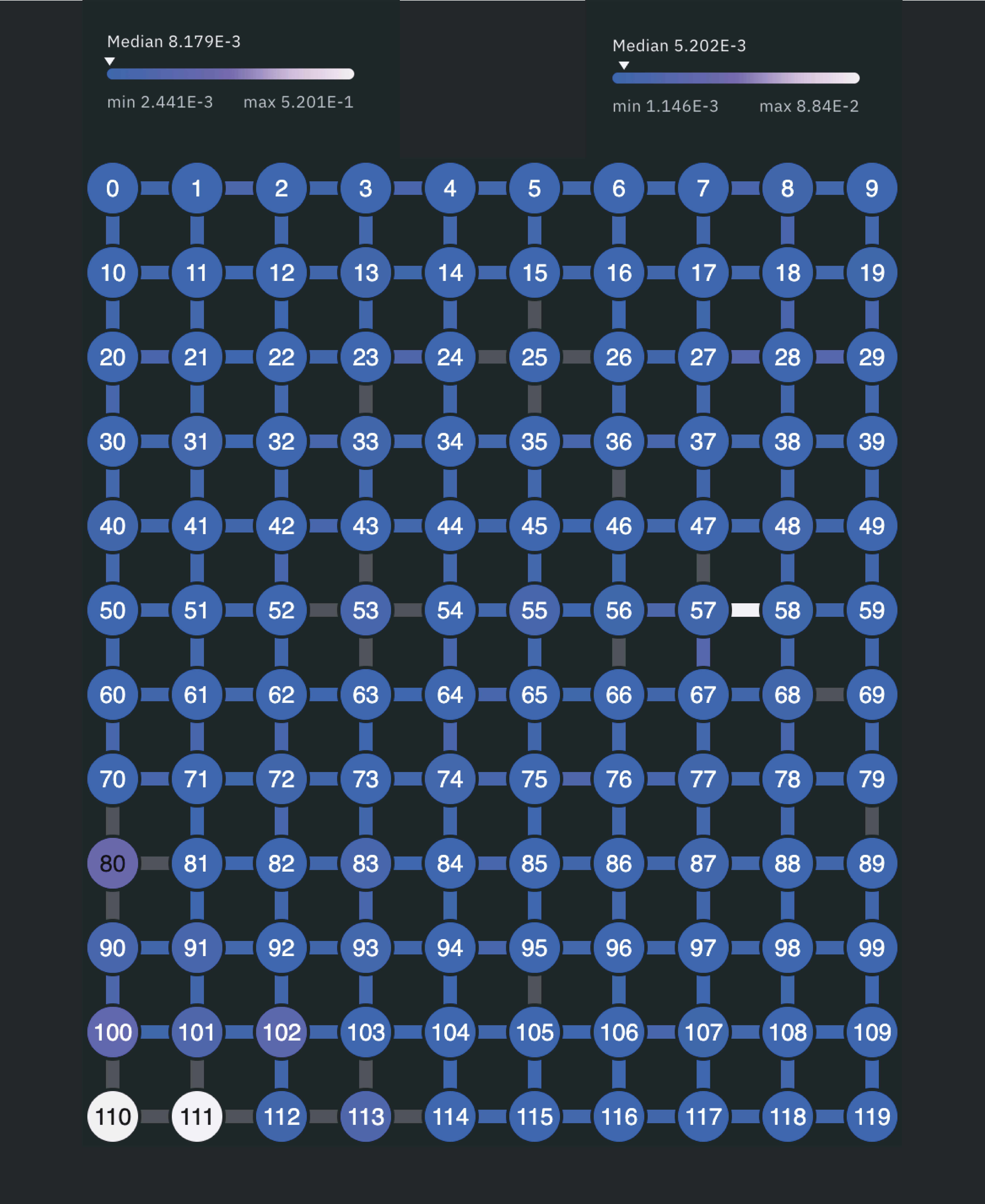}
\caption{Coupling map of \texttt{ibm\_rensselaer}. Nodes and their color represent physical qubits and their readout errors. Edges and their color represent available two qubit couplings and the corresponding controlled-$Z$ (CZ) gate errors.}
\label{fig:coupling_map}
\end{figure}

The quantum optimizations were executed on both the \texttt{ibm\_rensselaer} and \texttt{ibm\_miami} gate based quantum processor. Both systems use the IBM Nighthawk R1 chips, containing 120 qubits arranged in square lattice topology. Figure~\ref{fig:coupling_map} shows the coupling map of the \texttt{ibm\_rensselaer} quantum processor. The coupling map specifies which physical qubits are directly connected and can support native two qubit interactions, and it is represented graphically as a network in which nodes correspond to qubits and edges indicate available couplings between them.

\subsection{Experiment Setup}

The experiments were designed to compare our proposed QUBO with MIQUBO and with classical feature selection methods on the same three ASD datasets. Both QUBO formulations were solved using BF-DCQO and QAOA on the \texttt{ibm\_rensselaer} and \texttt{ibm\_miami} gate based quantum processors.

BF-DCQO was executed using five iterations and 1000 measurement shots per sampling step. For QAOA, the circuit depth was fixed at $p=2$, and the COBYLA optimization was limited to six objective function evaluations with 10,000 measurement shots per circuit evaluation. Six evaluations were used to provide sufficient objective evaluations for COBYLA to construct its initial local interpolation model and perform a subsequent parameter update. The larger shot count was selected based on preliminary experiments, in which 1000 shots produced greater variability in the QAOA solutions, whereas 10,000 shots produced more stable objective values and selected feature subsets. Repeated BF-DCQO executions produced the same feature subsets for the evaluated instances, whereas QAOA produced different selected subsets across runs. We therefore evaluated the QUBO formulation over 10 independent QAOA runs on Dataset 1 to characterize this variability. The same QUBO instances were also solved classically using ITS in MATLAB, providing a classical optimization reference for the quantum solutions.

The classical filter, wrapper, and embedded feature selection methods described above were independently applied to the same datasets. All classical and quantum methods returned five-feature subsets. The resulting subsets were evaluated using LDA, linear SVM, LR, RF, and PLS-DA, with classification accuracy measured using the same LOOCV procedure for all methods.

\section{Results}\label{sec3}

\subsection{QUBO}

\begin{figure*}[htbp]
\centering
\includegraphics[width=\textwidth]{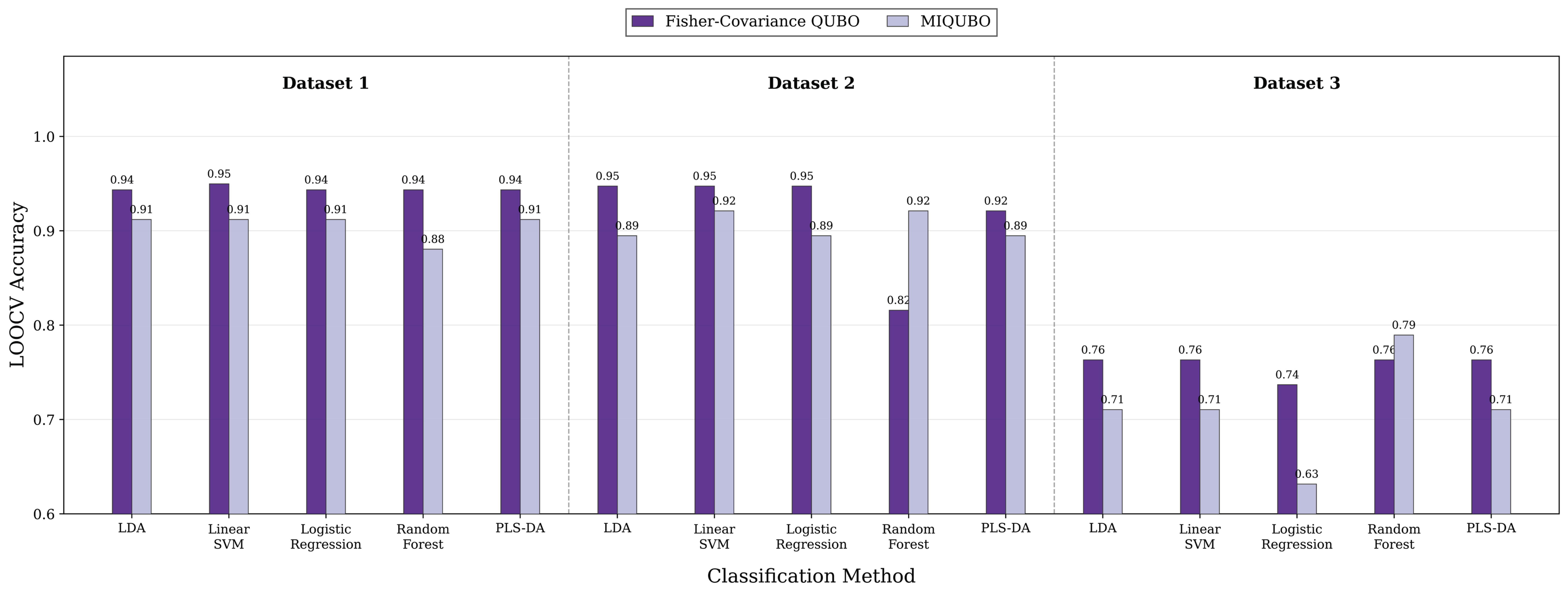}
\caption{LOOCV accuracy of LDA, linear SVM, LR, RF, and PLS-DA on Datasets 1, 2, and 3. The classifiers are trained on (1) 5 features selected by our proposed QUBO (Fisher-Covariance QUBO), and (2) 5 features selected by MIQUBO, both using BF-DCQO.}
\label{fig:qubo_comparison}
\end{figure*}

We compared the proposed QUBO with MIQUBO using BF-DCQO on all three datasets. Both QUBO formulations were additionally solved using QAOA on Dataset 1 to assess their behavior under a variational quantum optimization approach. QAOA was not evaluated on Datasets 2 and 3 because of its difficulty achieving stable convergence as the problem size increased due to the noise generated by greater circuit depth. In all tested cases, the QUBO solutions obtained using BF-DCQO were consistent with those obtained using ITS. The resulting five feature subsets were evaluated using LDA, linear SVM, LR, RF, and PLS-DA under LOOCV. The BF-DCQO results are shown in Fig.~\ref{fig:qubo_comparison}.

Solved using BF-DCQO, our proposed QUBO produced feature subsets that achieved higher classification accuracy than MIQUBO across most classifiers and datasets, with better performance in 13 of the 15 classifier--dataset comparisons. This indicates that the proposed QUBO is capable of identifying informative features that are useful for classification. The selected subsets also showed more stable performance across classifiers, particularly for Dataset~3, where the accuracies obtained using the proposed QUBO varied less than those obtained using MIQUBO. These results suggest that the Fisher score relevance and covariance redundancy formulation can generate feature subsets that are more discriminative and stable across different classification models than MIQUBO on the investigated metabolomic datasets.

\begin{figure}[htbp]
\centering
\includegraphics[width=\textwidth]{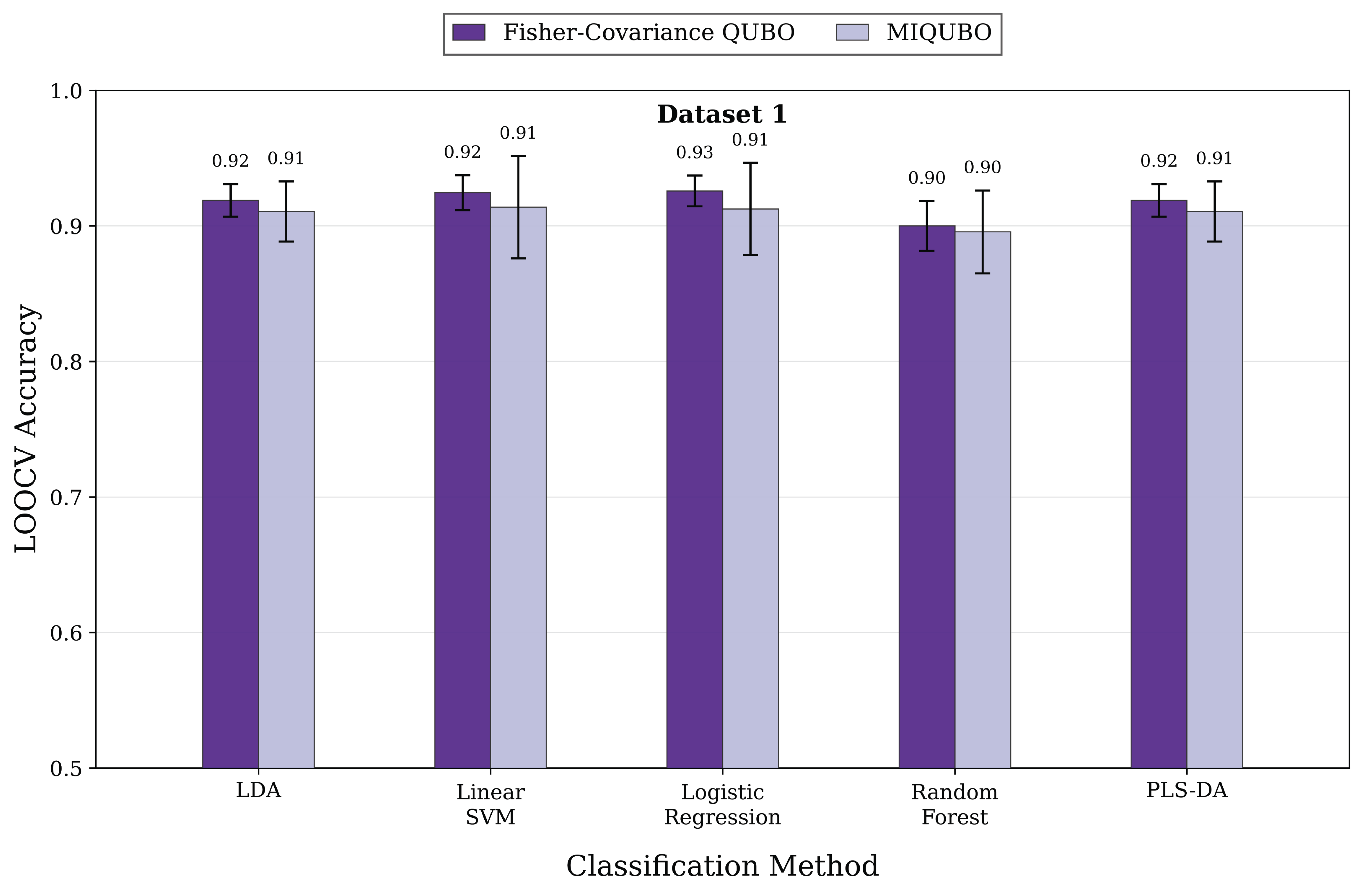}
\caption{LOOCV accuracy of LDA, linear SVM, LR, RF, and PLS-DA on Dataset 1 using 5 feature subsets selected by the proposed QUBO and MIQUBO with QAOA. Results are averaged over 10 independent QAOA runs for each QUBO formulation, with error bars representing one standard deviation.}
\label{fig:qaoa_qubo_comparison}
\end{figure}

The QAOA results on Dataset 1 showed a similar trend across repeated runs, as shown in Fig.~\ref{fig:qaoa_qubo_comparison}. Averaged over 10 independent QAOA runs, the proposed QUBO objective achieved higher average LOOCV accuracy than MIQUBO for all five classifiers. The proposed QUBO also produced smaller standard deviations across all classifiers, indicating more stable classification performance across repeated executions. These indicate that our formulation provides more robust feature selection outcomes under the stochastic nature of QAOA.

\subsection{Quantum and Classical Comparison}

\begin{figure*}[htbp]
\centering
\includegraphics[width=\textwidth]{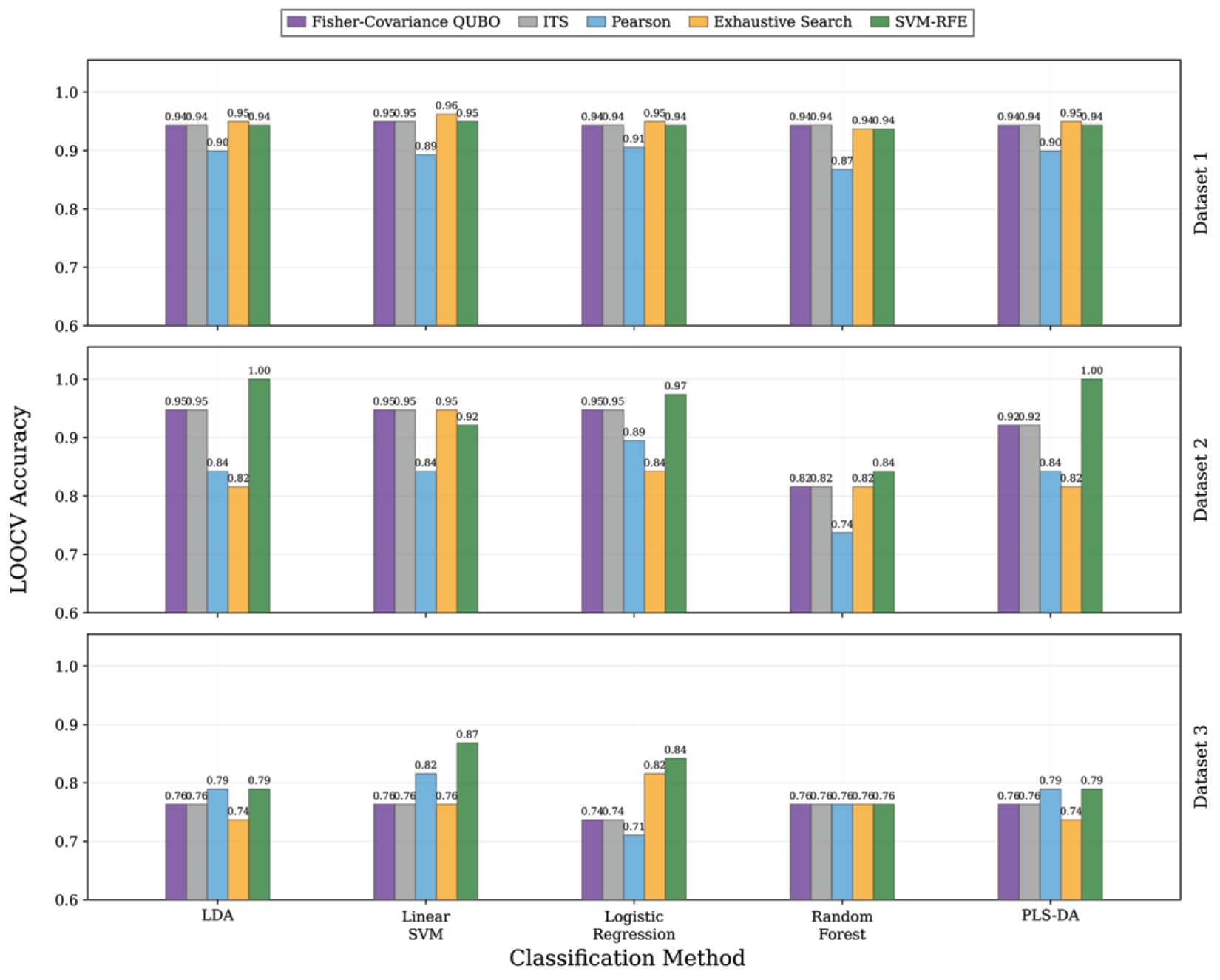}
\caption{LOOCV accuracy of LDA, linear SVM, LR, RF, and PLS-DA using five feature subsets selected by Fisher-Covariance QUBO, ITS, Pearson correlation, exhaustive search, and SVM-RFE on Datasets 1--3. The Fisher-Covariance QUBO was solved using BF-DCQO, and ITS was solved using MATLAB.}
\label{fig:quantum_classical_comparison}
\end{figure*}

The feature subsets selected by our proposed Fisher-Covariance QUBO were compared with those obtained using ITS, Pearson correlation, exhaustive search, and SVM-RFE. The Fisher-Covariance QUBO was solved using BF-DCQO, and ITS was used as a classical solver for the same QUBO formulation. For each method, the selected five feature subsets were evaluated using LDA, linear SVM, LR, RF, and PLS-DA under LOOCV. The results across the three datasets are shown in Fig.~\ref{fig:quantum_classical_comparison}.

Across all runs, BF-DCQO and ITS selected the same feature subsets and achieved identical LOOCV classification accuracies across all classifiers and datasets. This shows our proposed QUBO objective can produce consistent solutions regardless of its optimization methods. Overall, we observed that BF-DCQO selected feature subsets that were competitive with the classical feature selection methods. In most comparisons, the proposed QUBO achieved higher classification accuracy than the classical Pearson correlation and exhaustive search. The classification accuracies obtained from the QUBO solutions were also more stable across classifiers than those obtained using the classical feature selection methods, with smaller variation in performance across the evaluated classifiers. Across the three datasets, no single feature selection method consistently achieved the highest accuracy; instead, the relative performance depended on both the dataset and classification model.


\subsection{Runtime}

The average runtimes of BF-DCQO, ITS, and exhaustive search are summarized in Table~\ref{tab:runtime_comparison}. The problem sizes were 24, 56, and 54 features for Datasets 1, 2, and 3, respectively. For BF-DCQO, both the total workflow runtime and the QPU execution time are reported. A time to solution benchmark could not be performed here because the globally optimal solution was not independently known for the evaluated feature selection instances.

\begin{table}[htbp]
\centering
\caption{Average runtime in seconds for the quantum and classical optimization methods across datasets.}
\label{tab:runtime_comparison}
\begin{tabular}{lccccc}
\hline
Dataset & Features & BF-DCQO Total & BF-DCQO QPU & ITS & Exhaustive Search \\
\hline
Dataset 1 & 24 & 83.00 & 30.00 & 20.14 & 26.40 \\
Dataset 2 & 56 & 490.00 & 35.00 & 789.35 & 1741.38 \\
Dataset 3 & 54 & 383.50 & 35.00 & 1031.50 & 1463.93 \\
\hline
\end{tabular}
\end{table}

For the smaller 24 feature problem in Dataset 1, the classical methods required less runtime than BF-DCQO. For the larger 54 and 56 feature problems, however, BF-DCQO required less runtime than both ITS and exhaustive search. The reported BF-DCQO total runtime represents the complete runtime of workflow, including circuit transpilation, queue waiting time, and QPU execution. In contrast, the BF-DCQO QPU time represents only the time spent executing the quantum circuits on the quantum processor. The QPU execution time remained relatively consistent across the three problem sizes, ranging from 30s to 35s, suggesting that it was less affected by the increase in feature space size than classical methods. These results, however, do not establish a computational scaling advantage, since they are not time to solution comparisons. Assessment of quantum advantage requires benchmarking with clearly defined problem instances, computational resources, and performance metrics \citep{Koch2026}. Nevertheless, the results demonstrate the feasibility of quantum optimization for combinatorial feature selection and suggest its potential for larger feature spaces.

\section{Discussion}\label{sec4}

In this study, we proposed a novel QUBO objective for feature selection of biomedical metabolomics data. The formulation combines Fisher score as a measure of feature relevance with pairwise covariance as a measure of feature redundancy and can be solved using either classical or quantum optimization methods. Previous QUBO feature selection studies have demonstrated that representing feature importance and redundancy within a single binary optimization objective can provide competitive feature subsets and can be implemented using both classical and quantum solvers \citep{Muecke2023, Romero2025}. Our results extend this to a metabolomic dataset used for determinnig differences between children with an ASD diagnosis vs their typically developing peers. Across the three datasets, the proposed QUBO generally produced feature subsets with higher and more stable classification performance across classifiers than MIQUBO. When compared with representative classical filter, wrapper, and embedded methods, the QUBO solutions remained competitive across datasets and classification models. These results indicate that our proposed QUBO objective provides more robust feature selection solution than MIQUBO while maintaining classification performance that is competitive with classical feature selection methods.

One possible explanation for the stronger performance of our proposed QUBO over MIQUBO is the structure of the investigated data. Fisher score measures between-class separation relative to within-class variability and is suitable for data that are more linearly separable \citep{Gu2011}. The strong performance of LDA, linear SVM, and LR further suggests that the selected subsets contain class structure that can be captured by linear decision boundaries. The covariance term complements Fisher relevance by penalizing strongly co-varying features, which can reduce redundant information in pathway structured biomedical data. In contrast, MIQUBO uses mutual information for both relevance and redundancy \citep{Muecke2023}. Although mutual information can capture nonlinear dependencies, its estimation can be sensitive to limited sample size and the choice of discretization or estimator \citep{Paninski2003}. Fisher score and covariance rely on simpler sample statistics and may therefore provide more stable coefficients for the relatively small datasets considered here. Compared with classical methods, such as Pearson filtering, the proposed QUBO also accounts for pairwise relationships among selected features, while, unlike SVM-RFE, it is not tied to a specific predictive model. These properties may contribute to the more consistent performance observed across classifiers in our study.

We also evaluated the proposed QUBO using two gate based quantum optimization methods, BF-DCQO and QAOA, on IBM quantum hardware. We demonstrated our QUBO objective's compatibility with current NISQ devices, which are constrained by hardware noise, finite coherence, imperfect gates, limited connectivity, and circuit depth \citep{Preskill2018, Bharti2022, Cerezo2021}. These limitations are particularly relevant as larger feature selection problems require more pairwise interactions and deeper circuits. Previous hardware studies have also shown that QAOA performance can deteriorate with increasing problem size, limited hardware connectivity, and classical parameter optimization \citep{Harrigan2021, Zhou2020}. In our experiments, BF-DCQO produced the same solutions as ITS, while QAOA showed greater variability and required a deeper transpiled circuit. This difference reflects the differences in sampling and optimization structures of the two algorithms. BF-DCQO is a sample based iterative solver in which measured bitstrings are used to update the bias fields and concentrate the sampling distribution toward low energy solutions. Its performance therefore depends on obtaining a sufficient number of low error samples that contain high quality solutions. In contrast, QAOA combines a variational training stage with a final sampling stage. QAOA quantum measurements are required repeatedly during the variational loop to estimate the objective function and guide the classical parameter optimizer, followed by additional measurements of the optimized circuit to identify candidate solutions. Consequently, QAOA requires measurement resources during both parameter optimization and final solution sampling, where BF-DCQO uses measurements directly within its iterative sampling procedure. Hardware errors and increased circuit depth can further reduce the probability of obtaining useful samples, which may contribute to the greater variability observed for QAOA in this study. BF-DCQO also required less total runtime than ITS and exhaustive search for the 54 and 56 feature problems. However, these runtime results should not be interpreted as evidence of quantum advantage \citep{Koch2026}, because current quantum hardware constraints limited our experiments to relatively small or dimensionally reduced datasets. Nevertheless, the results indicate the feasibility of QUBO feature selection on current quantum processors and its potential for larger combinatorial search problems as quantum hardware continues to improve.

\begin{figure*}[htbp]
\centering
\includegraphics[width=\textwidth]{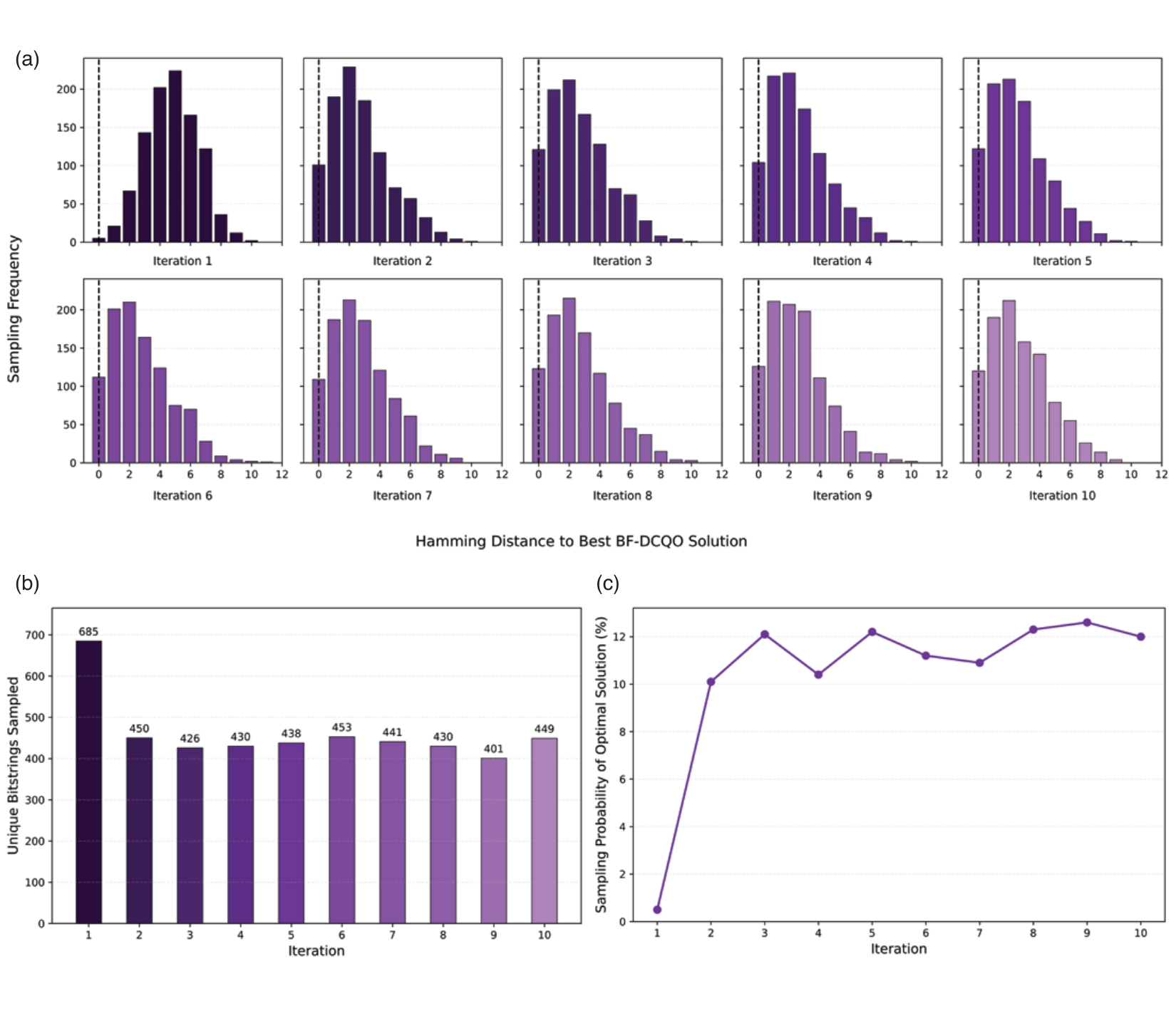}
\caption{Convergence of BF-DCQO on a reduced 12 features dataset demonstrated using 10 iterations and 1000 measurement shots per iteration.
(a) Sampling frequency distributions as a function of Hamming distance from the best observed solution across 10 iterations.
(b) Sampling diversity measured by the number of unique bitstrings observed in each iteration.
(c) Sampling probability of the best observed solution at each iteration.
The demonstration was performed separately from the primary Dataset 1 experiment, which used 24 features, five BF-DCQO iterations, and 1000 shots per sampling step.}
\label{fig:bfdcqo_convergence}
\end{figure*}

To better examine the convergence behavior of quantum feature selection method, we performed a separate demonstration using BF-DCQO on a reduced problem size of 12 feature with 10 iterations and 1000 measurement shots per iteration, as shown in Fig.~\ref{fig:bfdcqo_convergence}. This demonstration differs from the primary Dataset 1 experiment, which used 24 features, five iterations, and 1000 shots per sampling step, and was included specifically to illustrate the iterative convergence behavior of BF-DCQO over a longer sequence of iterations. In the first iteration, the sampled solutions were broadly distributed and concentrated at greater Hamming distances from the best observed solution. After the first bias-field update, the distribution shifted toward smaller Hamming distances and remained concentrated near the best solution in subsequent iterations--Fig.~\ref{fig:bfdcqo_convergence}(a). Consistent with this behavior, the number of unique sampled bitstrings decreased from 685 in the first iteration to approximately 400--450 in later iterations, indicating reduced sampling diversity and increasing concentration within a smaller region of the solution space--Fig.~\ref{fig:bfdcqo_convergence}(b). The sampling probability of the best observed solution also increased from below 1\% in the first iteration to approximately 10--13\% in subsequent iterations--Fig.~\ref{fig:bfdcqo_convergence}(c). Together, these results illustrate how the iterative bias-field updates in BF-DCQO progressively concentrate the sampling distribution around promising solutions.

Despite promising results, several limitations remain. First, our proposed QUBO objective calculates feature importance through linear terms and pairwise interactions through quadratic terms, while biological pathway data may involve higher order interactions among metabolites, genes, or microbial features. Higher-Order Unconstrained Binary Optimization (HUBO) provides a potential extension by incorporating third or higher order interactions. Recent studies have demonstrated higher order quantum feature selection on trapped-ion hardware \citep{FloresGarrigos2026}, higher order BF-DCQO on IBM quantum hardware \citep{RomeroHUBO2025}, and QAOA for higher order binary optimization \citep{Kiktenko2026}. However, the number of feature interaction terms and required circuit complexity increase with feature dimensionality, making evaluation of HUBO objective challenging on current NISQ hardware. More generally, hardware noise, limited connectivity, and executable circuit depth restricted this study to smaller or reduced dimensional datasets \citep{Proctor2022}. As a result, Datasets 2 and 3 required dimensionality reduction before our evaluation, and QAOA was evaluated only on Dataset 1 because of increasing circuit complexity at larger problem sizes. Future improvements in qubit quality, connectivity, coherence, and executable circuit depth may enable the proposed framework to be extended to higher order interactions, larger datasets, and higher dimensional feature spaces.

\backmatter

\bmhead{Acknowledgments}
The authors thank Michael Sofka and the Quantum Enablement and Development (QED) group at Rensselaer Polytechnic Institute for their support in assistance with quantum hardware logistics.

\section*{Declarations}

\bmhead{Funding}
No funds or grants were received.

\bmhead{Competing interests}
R.L. is an employee of Kipu Quantum GmbH and holds a financial interest in the company, which develops the BF-DCQO solver evaluated in this study. All experiments for both BF-DCQO and QAOA were implemented and executed by H.L. and B.C., who have no affiliation with or financial interest in Kipu Quantum GmbH. H.L., B.C., S.B., J.H., Z.L., and Z.Z. declare no competing interests.

\bmhead{Ethics approval and consent to participate}
Not applicable. Clinical trial number: not applicable.

\bmhead{Consent for publication}
All authors have read and approved the final manuscript and consent to its submission for publication.

\bmhead{Data availability}
The datasets are publicly available at \url{https://github.com/Liu-Machine-Learning/quantum-feature-selection.git}.

\bmhead{Materials availability}
Not applicable

\bmhead{Code availability}
BF-DCQO is a quantum optimization algorithm developed by Kipu Quantum GmbH and described in previous studies \citep{Hegade2022,GomezCadavid2025}. The exact implementation of BF-DCQO and the orchestration of its pipeline stages are proprietary to Kipu Quantum GmbH and are not publicly available. The software was obtained by the RPI authors under a commercial licence and accessed through the Kipu Quantum Hub \url{https://hub.kipu-quantum.com}. Requests for access should be made directly to Kipu Quantum GmbH. The QAOA implementation, problem instances, and all analysis code used in this study are available at \url{https://github.com/Liu-Machine-Learning/quantum-feature-selection.git}.

\bmhead{Author contribution}
H.L., J.H., and Z.Z. conceptualized the study. H.L. and R.L. developed the methodology. B.C. and H.L. performed the formal analysis and investigation. H.L. prepared the original manuscript draft. B.C., H.L., J.H., R.L., S.B., Z.L., and Z.Z. reviewed and edited the manuscript. J.H. supervised the study. All authors reviewed and approved the final manuscript.

\bibliography{sn-bibliography}

\end{document}